\documentclass{emulateapj}
\begin{document}
\title{Evolution of Rotating Accreting White Dwarfs and the Diversity of Type I\lowercase{a} Supernovae}
\author{Tatsuhiro Uenishi, Ken'ichi Nomoto}
\affil{Department of Astronomy, School of Science, University of Tokyo}
\and
\author{Izumi Hachisu}
\affil{Department of Earth Science and Astronomy,\\
College of Arts and Science, University of Tokyo}

\slugcomment{To appear in the October 1 2003 issue of the Astrophysical Journal (v595n2) from pages 1094-1100.}

\begin{abstract}
Type Ia supernovae (SNe Ia) have relatively uniform light curves and spectral evolution, which make SNe Ia useful standard candles to determine cosmological parameters.  However, the peak brightness is not completely uniform, and the origin of the diversity has not been clear. We examine whether the rotation of progenitor white dwarfs (WDs) can be the important source of the diversity of the brightness of SNe Ia. We calculate the structure of rotating WDs with an axisymmetric hydrostatic code. The diversity of the mass induced by the rotation is $\sim 0.08 M_{\odot}$ and is not enough to explain the diversity of luminosity. However, we found the following relation between the initial mass of the WDs and their final state; i.e., a WD of smaller initial mass will rotate more rapidly before the supernova explosion than that of larger initial mass. This result might explain the dependence of SNe Ia on their host galaxies.
\end{abstract}


\section{INTRODUCTION}
Type Ia supernovae (SNe Ia) are widely believed to be thermonuclear explosions of mass-accreting white dwarfs (WDs), whose masses have reached a certain critical level (e.g., \citet{Nomoto97, Branch98} for reviews). However, the observed brightnesses of SNe Ia have a little diversity. SNe Ia have been standardized by applying the empirical relation existing between the peak brightness and the light curve width which is obtained from nearby SNe Ia \citep{Phil93,Hamuy96,RPK96}. Explaining this diversity is very important to the use of SNe Ia as standard candles for the determination of the cosmological parameters because whether the relation is valid for high-redshift SNe Ia is still uncertain. The diversity is likely to be related to the evolution of the progenitor systems \citep{Branch95,Livio00,Nomoto00}.

Theoretically, the critical mass for the explosion of SNe Ia depends on their progenitor models. There are two groups of models according to the masses of exploding WDs: Chandrasekhar-mass models and sub-Chandrasekhar-mass models. In the Chandrasekhar-mass models, a WD is supposed to explode at the center when its mass becomes very close to or exceeds the Chandrasekhar's limiting mass ($\sim 1.4 M_{\odot}$). Because of presence of a limiting mass, the property of the WDs is almost identical in the Chandrasekar-mass model. This property would lead to the almost uniform peak brightnesses of SNe Ia. In the sub-Chandrasekhar-mass models, on the other hand, an explosion is initiated at the bottom of the helium layer before the WD mass reaches the Chandrasekhar limiting mass \citep{Woosley94}. Inward deflaglation waves converge at the center of the WD and ignite detonation waves, which blow off the whole WD \citep{Woosley86}. Sub-Chandrasekhar-mass models have spectra that are too blue and are not in good agreement with the observed spectra of standard SNe Ia \citep{Hoeflich96,Nugent97}. However, some of the peculiar SNe Ia, such as SN 2000cx have bluer spectra than standard ones \citep{Li01} and may be explained by the sub-Chandrasekhar-mass models.

Two possible scenarios have been considered for the Chandrasekhar-mass models. One is the double degenerate (DD) scenario in which two WDs in a binary system merge together as gravitational radiation extracts the orbital energy from the system \citep{Iben84, Webbink84}. As a result, the two WDs turns into one WD and a thick disk surrounding it whose total mass exceeds the Chandrasekhar mass. However, \citet{SN85, SN98} shows that the merged WDs cannot result in SNe Ia because of the relatively high accretion rate from a thick disk. They are likely to turn into O+Ne+Mg WDs, then finally become neutron stars via accretion-induced collapse. The other is the single degenerate (SD) scenario in which a C+O WD in a close binary system accretes hydrogen-rich matter from its companion star to grow to the Chandrasekhar mass. Such systems are actually observed as supersoft X-ray sources. Therefore, Chandrasekar mass model in the SD scenario is the most favored model now for standard Sne Ia.

The evolution of accreting WDs have been extensively investigated by many authors \citep{Nomoto82a,Kobayashi98,Kobayashi00,L&v97,Langer00}. To make a supernova explosion, a WD needs to get enough mass from its companion star. When a star becomes a WD in a binary system, its companion star is still on the main sequence. As the companion evolves, its envelope expands and fills its Roche lobe. Then gas flow toward the WD starts and forms an accretion disk. The WD begins to gain the mass and angular momentum from the rotating disk.

The evolution of a mass-accreting WD depends mainly on the accretion rate ($\dot{M}$). A WD needs an appropriate accretion rate to become a SN Ia \citep{Nomoto82b}. Too low $\dot{M}$ ($\dot{M} < 10^{-8}M_{\odot} \textrm{yr}^{-1}$) causes a surface nova explosion and the WD loses its mass. Too high $\dot{M}$ causes the WD to expland to the size of a red giant and leads to the formation of a common-envelope. After the common envelope forms, the gas that has overflowed the Roche robe of the system reduces the angular momentum of the system, which induces merging of the two stars \citep{Nomoto79}. It has been thought that a SN Ia is not expected to occur for such very low $\dot{M}$ or very high $\dot{M}$.

However, the strong wind solution is found for $\dot{M}$ over the certain critical value using updated OPAL opacity. If $\dot{M} > \dot{M}_{\textrm{wind}}$ the WD blows strong winds to avoid the common-envelope formation (see \citet{Hachisu96,Hachisu99a,Hachisu99b} for details of the wind model). Therefore, more WDs can accrete mass from the companion stars until the mass reaches the critical mass for SNe Ia.

Previous studies on the SNe Ia progenitors have ignored rotation or used averaged centrifugal force assuming spherical symmetry (e.g., \citet{Langer00,Langer03,Pier03}). However, it is natural to think that accreting WDs are rotating because they must gain not only mass but also angular momentum from the accretion disk. If rotation of the accreting WD is fast enough, the shape of the WD is distorted and the difference from the spherical case cannot be ignored (e.g., \citet{Wang03}).

Although it is very important to consider rotation of the progenitors properly, there has been little systematic study on the multi-dimensional structure of rotating WDs in connection with the brightness of SNe Ia. Our main objective are to investigate the strucuture and evolution of accreting WDs in close binaries with a two-dimensional code and to see (1) to what extent the rotation of the WD affect the strucuture and (2)to what extent the diversity of SNe Ia due to the progenitor's rotation can be obtained.

An accreting WD gains angular momentum from the rotating disk as well as mass, and gets rotating faster and faster. Rapid rotation affects the mass and structure of a WD just before the supernova explosion. The existence of centrifugal force effectively reduces gravity and thus it is possible to sustain a larger amount of matter with the same degenerate pressure. Thus, the critical mass will increase as rotation becomes faster.

When the rotating WD reaches the critical mass (to some extent larger than that of nonrotating WDs), it will explode as a SN Ia. The increase of mass and the change of inner structure will affect its nucleosynthesis. The brightness of a SN Ia is mainly dependent on the amount of $^{56}$Ni synthesized in the explosion; therefore rotation of progenitor WDs will affect the brightness, too.

In \S 2, we describe models and assumptions. The results of our calculations are shown in \S 3. Disucussion is in \S 4 and the conclusions are summerized in \S 5.

\section{MODELS}

\subsection{Basic Assumptions}

To calculate the structure of rotating WDs, we use Hachisu's axisymmetric code (Hachisu Self-Consistent Field [HSCF] method: \citet{Hachisu86}). The number of meshes are 129 in both radial and angular directions.

We adopt the equation of state of complete degenerate electron gas for whole WD.

\begin{eqnarray}
P&=&A\left[ x(2x^2 -3)(x^2 + 1)^{1/2} + 3\ln (\sqrt{1+x^2}+x)\right]\\
\rho &=& Bx^3
\end{eqnarray}
where
\begin{eqnarray}
x \equiv \frac{p_F}{mc}, \qquad p_F = \left( \frac{3h^3}{8\pi}n_e \right)^{\frac{1}{3}}, \qquad n_e = \frac{N_0 \rho}{\mu_e}\\
A=6.002 \times 10^{22} \mathrm{dynes \cdot cm^{-2}}\\
B=9.736 \times 10^5 \mu_e \mathrm{g \cdot cm^{-3}},
\end{eqnarray}
and $\mu_e$ is the mean molecular weight of electrons and fixed as $\mu_e = 2$ in all calculations.; $N_0$ is the Avogadro's number, $n_e$ is the mean electron number density and $p_F$ is the Fermi momentum. No Coulomb interaction is included.

We assume that a WD is at rest when it starts to gain mass and angular momentum; i.e. $J=\Omega=0$ at $M=M_{\mathrm{int}}$, where $M_{\mathrm{int}}$ is the initial mass of the WD. Angular momentum of accreting gas is supposed to obey Keplerian rotation law, and gives all of its angular momentum to the WD. Under these assumptions, specific angular momentum $j$ is given as follows:
\begin{eqnarray}
j = j_{\mathrm{Kepler}} = R \sqrt{\Phi_{\mathrm{surface}}} \label{j}, \label{alpha}\\
\Phi_{\mathrm{surface}} = -G\int{\left[ \frac{\rho(\mathbf{r}')}{|\mathbf{r}-\mathbf{r}'|} d^3\mathbf{r}'\right]},
\end{eqnarray}
where $R$ is the equatorial radius of a WD and $\Phi_{\mathrm{surface}}$ is its surface gravitational potential. At each step of the calculation, the WD is kept in hydrostatic equilibrium by the balance between the gravitational force, centrifugal force, and the pressure gradient.
\begin{equation}
\int{\rho^{-1}dP}+\Phi-\int{\Omega^2wdw}=C(\mathrm{constant}),
\end{equation}
where $\Omega$ is the angular velosity and $w$ is the distance from rotation axis; i.e., $w=r(1-\mu^2)^{1/2}$ in spherical polar coordinates $(r,\mu,\phi)$.

The mechanism and timescale of the angular momentum transfer in the star have not been well understood. We thus adopt the assumption that the WD is separated into two regions. One is the central, nonrotating region, and the other is surrounding rigidly rotating region, although in reality there exists a gradient of the angular velocity \citep{Yoon02, Yoon03}. The ratio of the mass of the former region to the whole WD is denoted as,
\begin{equation}
	\xi = \frac{M_{\mathrm{NR}}}{M},
\end{equation}
where $M_{\mathrm{NR}}$ is the mass of the nonrotating region. A WD is initially at rest, i.e., $\xi_0 = 1$. We consider two extreme cases of anglar momentum transfer: (1)Angular momentum transfer is fast and the whole WD is rotating rigidly ($\xi = 0$). (2)Angular momentum transfer is slow and initially nonrotating WD remains static while the accreting gas is rigidly rotating around it ($\xi = M_{\mathrm{int}}/M$).

In each case, the angular velocity $\Omega$ is expressed as follows:
\begin{eqnarray}
	\Omega &=& \Omega_0(=\mathrm{constant}) \qquad (M(w) > M_{\mathrm{NR}}) \label{Eq:rot.a}\\
	\Omega &=& 0 \qquad (M(w) < M_{\mathrm{NR}})\label{Eq:rot.b}
\end{eqnarray}
where $M(w)$ is the mass included in radius $w$.

The HSCF method is a variation of the self-consistent method to solve the axisymmetric stellar structure and characterized by the selection of parameters \citep{Hachisu86}. To obtain the structure of a WD, one should specify its central density ($\rho_{\mathrm{c}}$) and the axis ratio ($q$) in the HSCF method. The axis ratio ($q$) of the WD is the ratio of its polar radius to its equatorial radius, which is an index of the strength of rotation. If a WD rotates rapidly, its shape is distorted by the centrifugal force and its polar radius becomes shorter than its equatorial radius. For a spherical WD, $q=1$, while, a WD with higher angular velocity has smaller $q$.

In this calculataion we need another parameter $\xi$ (the ratio of the mass of the nonrotating region to the whole WD; eq \ref{Eq:rot.a}, \ref{Eq:rot.b}) to specify the structure. We have calculated more than 100,000 WD models with different sets of $(\rho_{\mathrm{c}}, q, \xi)$. For each $\xi$, the ranges of $\rho_\mathrm{c}$ and $q$ are bounded by the criteria of the supernova explosion and the critical rotation. Each criterion is described in \S 2.2.

\subsection{Criteria}
The WD is compressed as its mass increases. On the other hand, rotation mitigates compression by reducing effective gravity. For given $J$, the maximum density $\rho_{\mathrm{max}}$ is higher for larger $M$. For given $M$, $\rho_{\mathrm{max}}$ is higher for smaller $J$. A thermonuclear explosion is expected to occur when the $\rho_{\mathrm{max}}$ exceeds a certain critical value. Although the actual ignition density depends on the temperature (and thus $\dot{M}$), geometry and the explosion models, we adopt $2\times10^9 \mathrm{gcm}^{-3}$ for the ignition density as in the one-dimensional model W7 \citep{Nomoto84} in the present approach with the complete degenerate model sequences.

The criterion for the critical rotation of a rotating WD depends on the rotation law. In rigid rotation, which we adopt in the present study, it is given as
\begin{equation}
\eta = \frac{R^3 \Omega ^2}{GM} = 1,
\end{equation}
where $\eta$ is the ratio of the equatorial centrifugal force to the gravity. A WD could break-up at equatorial plane when $\eta$ becomes unity.

\begin{figure*}[hb]
\plottwo{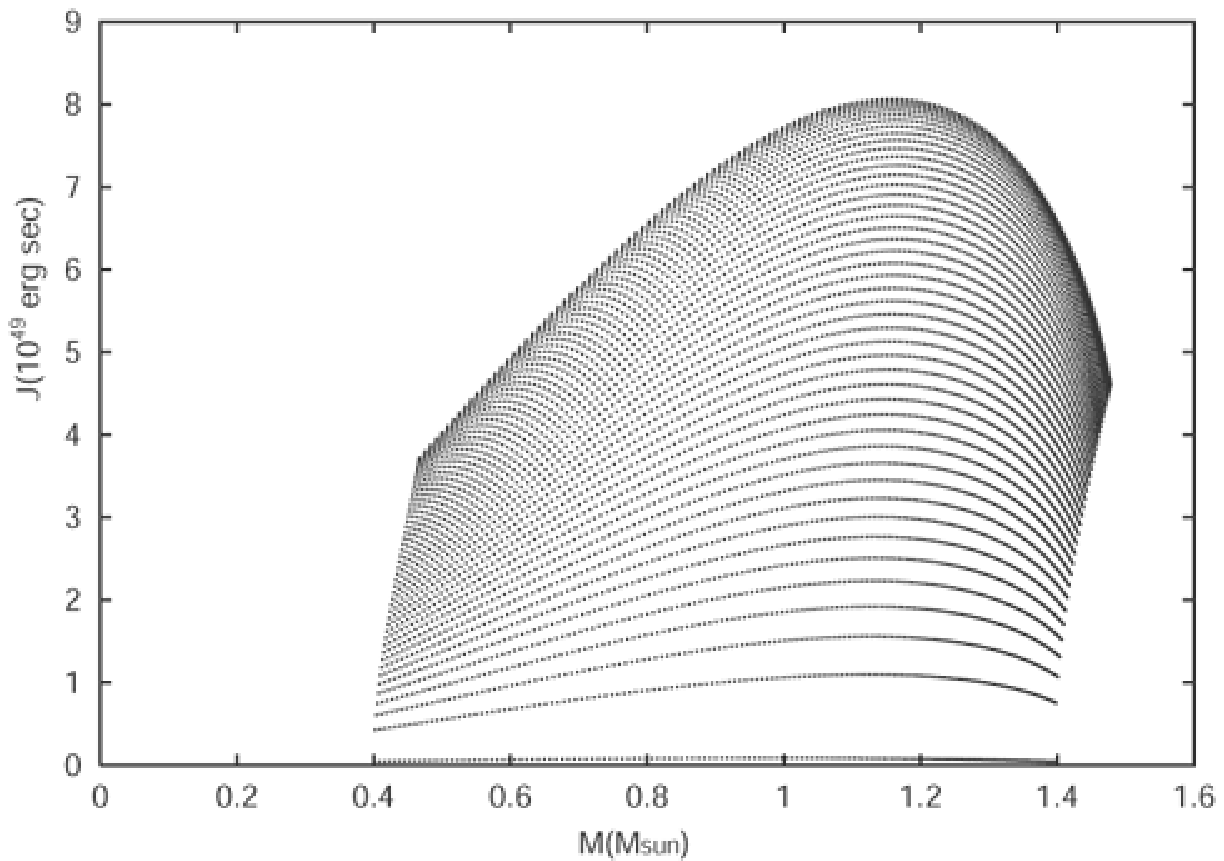}{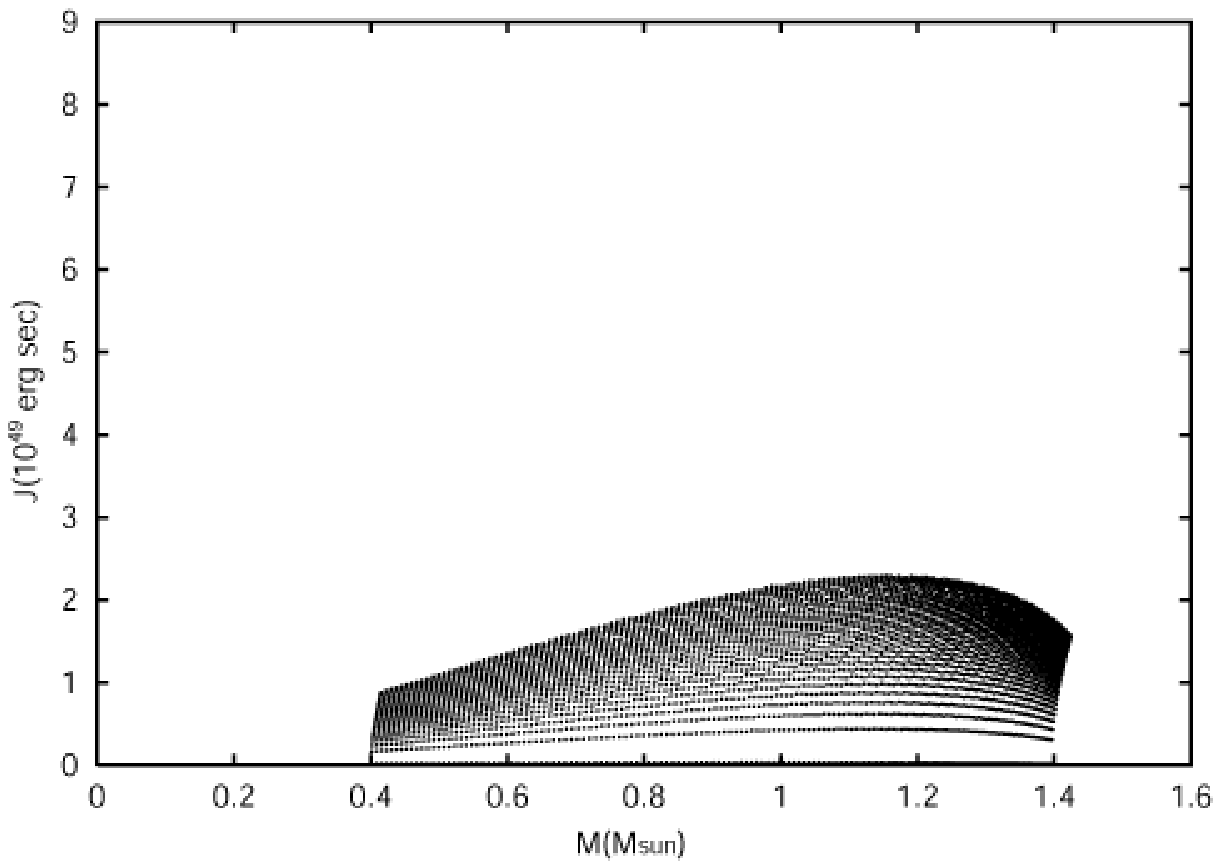}
\caption{The total data area in $M-J$ plane where WDs can exist stably under the assumptions mensioned in the text. The left panel is for the case of true rigid rotation ($\xi = 0$), and the right panel is for the case of slow angular momentum transfer ($\xi = 0.9$), respectively.}
\label{Figure:MJ}
\end{figure*}

Figure \ref{Figure:MJ} show the area in the $M-J$ plane where WDs can exist stably under these criteria. Each point represents a WD of a different set of $(\rho_{\mathrm{max}},\eta)$, which is transformed into (M,J). The sequence of WD models with $\rho_{\mathrm{max}} = 2.0 \times 10^9$ g cm$^{-3}$ forms right edge of data area, and upper edge is composed of WDs whose ratio of surface gravity to centrifugal force are unity ($\eta = 1$). As seen in Figure \ref{Figure:MJ}, WDs of lower $\xi$ (i.e., with a smaller non-rotating region) have wider stable region in the parameter space. For $\xi = 1$, all WDs are on the M-axis (no-rotation spherical case).

\begin{figure*}[ht]
\plottwo{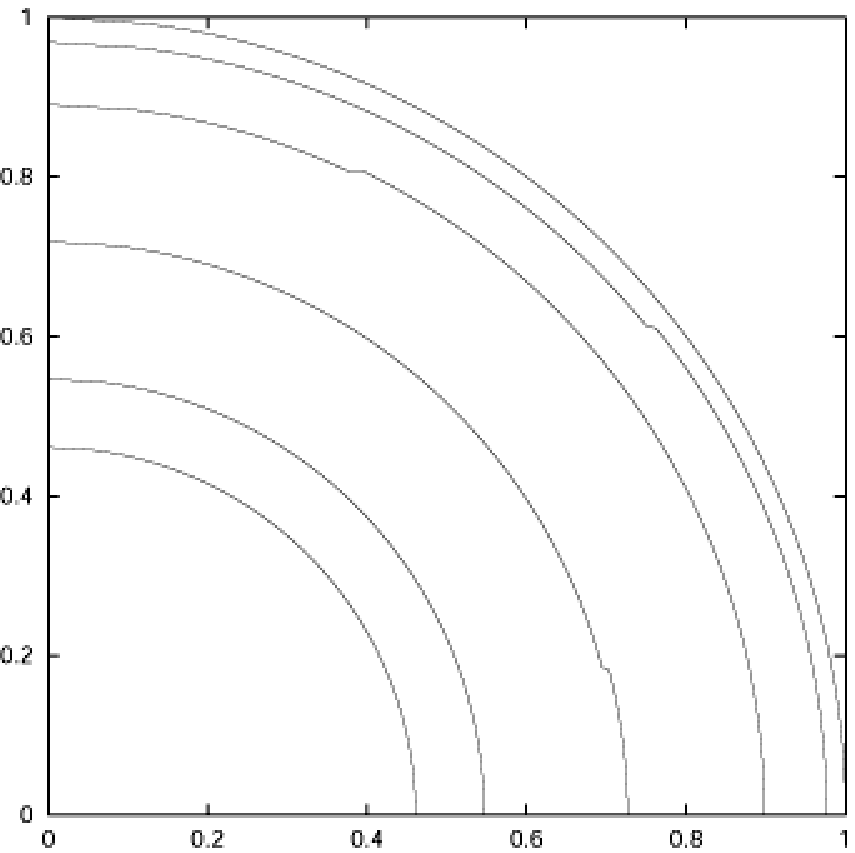}{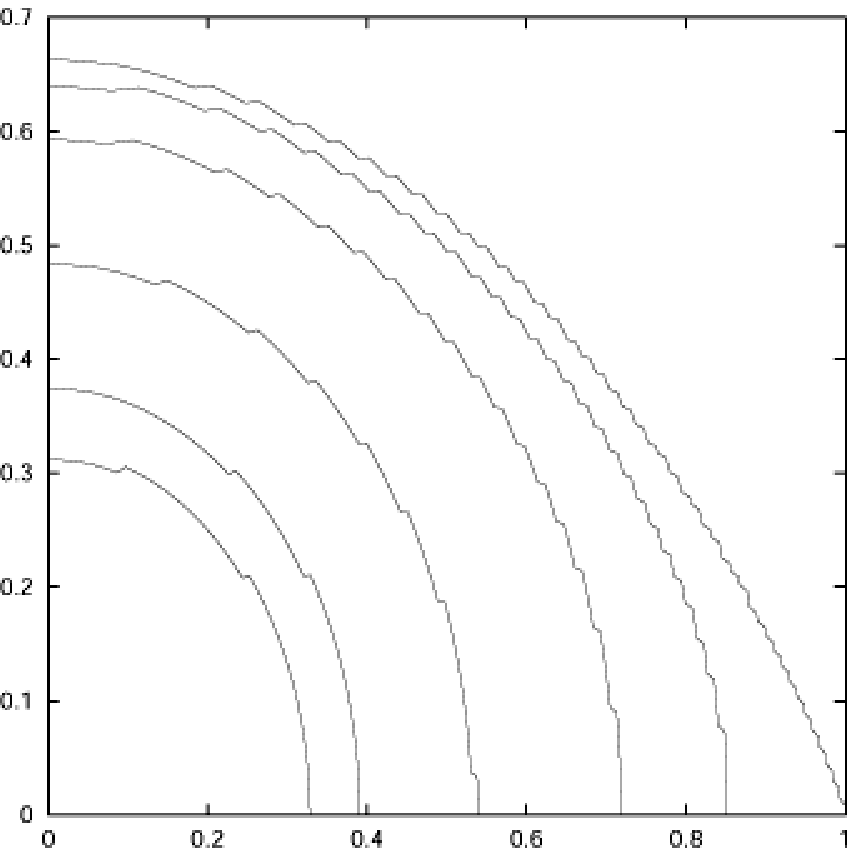}
\caption{Density contour in the meridional cross section for $\rho_c = 2.0 \times 10^9$ (just before the supernova explosion). Rotation law is true rigid rotation ($\xi = 0$). The left panel is nearly spherical case ($\eta \sim 0$) and the right panel is nearly critical rotation case ($\eta \sim 1$).}
\label{Figure:Contour.Rigid}
\end{figure*}

\begin{figure*}[hb]
\plottwo{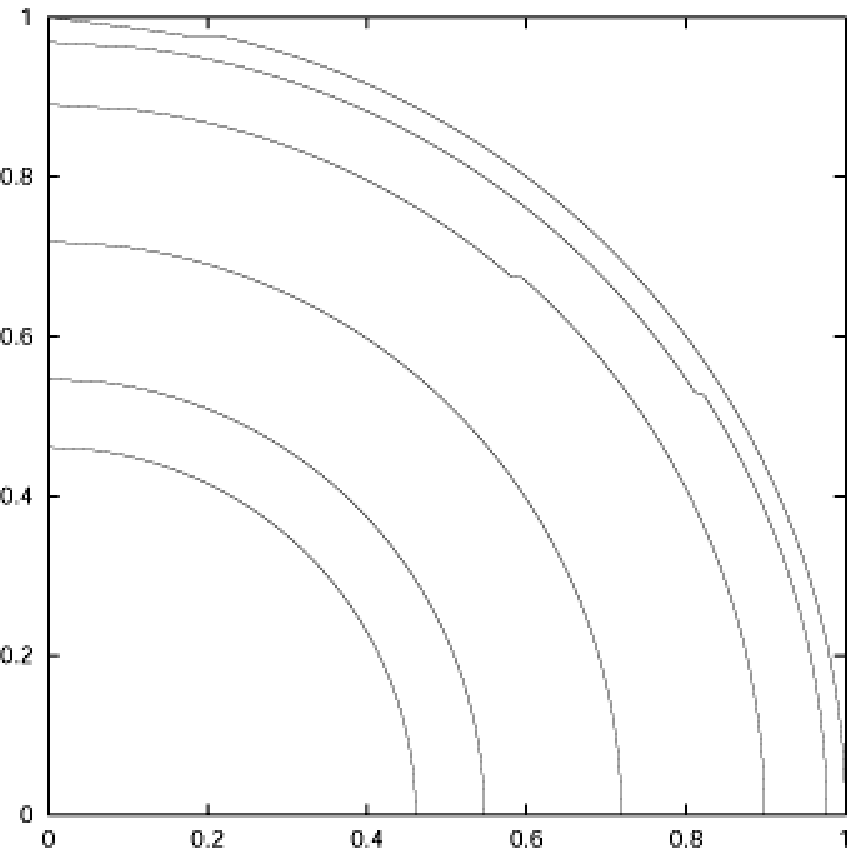}{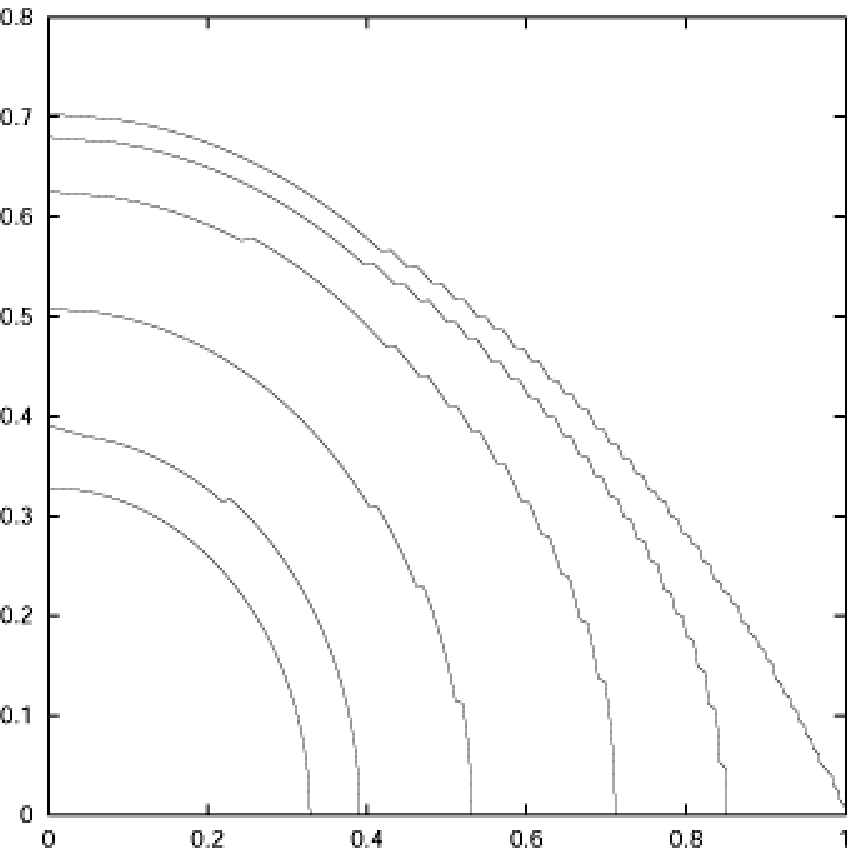}
\caption{Same as Figure 2, but for $\xi = 0.9$.}
\label{Figure:Contour.xi09}
\end{figure*}

In Figures \ref{Figure:Contour.Rigid} and \ref{Figure:Contour.xi09}, examples of the structure of the accreting WDs just before the supernova explosion (i.e, $\rho_c = 2.0 \times 10^9$ g/cm$^3$) are shown. The equatorial radius and polar radius are normalized by the equatorial radius. In both figures, the left panels are almost nonrotating case ($\eta \sim 0$, axis ratio $q = 128/129$; maximum value of $q$ for rotating WDs in this method) and the right panels are almost break-up case ($\eta \sim 1$).

What would occur when a WD reaches the upper edge of the stable area by accretion? Is the WD broken up? Does it trigger the mass shedding and divides itself into several pieces? There is one possible answer for this question. \citet{Pac90} and \citet{P&N90} found a solution that permits the WD to accrete without becoming unstable or triggering mass shedding, by computing a system consisting of a rigidly rotating WD and its viscous accretion disk as one fluid. (see also \citet{Fujimoto95}).

Normaly, a rigidly rotating WD gains angular momentum from the Keplerian accretion disk due to viscosity. Inward angular momentum flow spins up the WD to increase $\Omega_0$. When $\Omega_0$ exceeds the critical value ($\Omega_{\textrm{crit}}$) and $\eta$ exceeds unity, angular momentum is transported backwards from the WD to the disk while the mass of the WD increases. This state of rotation is called "supercritical rotation" \citep{Pac90}. Deprived of its angular momentum, the WD slows down and when its angular velocity falls below $\Omega_{\textrm{crit}}$, the supercritical rotation ends and the WD gains both mass and angular momentum again. The viscous accretion disk acts like a stabilizer under the supercritical rotation. Then, if a WD reaches the critical rotation, we regard it will evolve along the upper envelope in the $J$-$M$ plane, and eventually explode as a SN Ia.

\subsection{Accretion Model}
In our accretion model, the initial state is defined only by the initial mass ($M_{\mathrm{int}}$). We consider $0.60 M_{\odot} < M_{\mathrm{int}} < 1.07 M_{\odot}$ for the initial mass range of a WD. This is the range where C+O WDs in the close binaries can form \citep[e.g,]{Umeda99b}. Therefore, the range of $\xi$ is $0.42 (=0.60/1.40) < \xi < 1.0$ in the slow angular momentum transfer. 

The accreting WD gains mass $\Delta M$ and angular momentum $\Delta J$ at each step. The value of $\Delta M$ is fixed throughout the sequence. The new mass $M_{\mathrm{new}}$ and angular momentum $J_{\mathrm{new}}$ are expressed as
\begin{eqnarray}
M_{\mathrm{new}} = M + \Delta M \label{mnew}\\
J_{\mathrm{new}} = J + \Delta J = J + j \Delta M \label{jnew}
\end{eqnarray}
where $j$ is specific angular momentum of accreting gas as given by eq.(\ref{j}).

When the new mass and angular momentum are obtained, the nearest data point in the stable area is searched. Next, two dimension interpolation is made using the surrounding four points in $M-J$ plane because we cannot directly caluculate the WD that have the exact value of $M_{\mathrm{new}}$ and $J_{\mathrm{new}}$. The obtained new potential and the specific angular momentum $j$ are used for the calculation of the next step of accretion.

In the slow-transfer case, the value of $\xi$ decreases as a WD accretes matter and its whole mass increases. Because we only have finite resolution in $\xi$ plane, we cannot calculate the model of exact value of $\xi$ in each step. Therefore, we use the data sets of WDs of the nearest $\xi$, and switch them as $\xi$ decreases. The area in $M-J$ plane where WDs can exist stably without the critical rotation depends on $\xi$, and WDs sometimes reach the critical rotation before $\xi$ becomes low enough to switch the data sets because of the low resolution in $\xi$ plane. We consider WDs evolve along the upper edge of each data sets same as the fast-transfer case until $\xi$ becomes low enough to use the next data set of $\xi$.

\section{RESULTS}

Many WD models are calculated for various parameters. For each initial mass, the evolution of WD is followed.

\subsection{Fast Angular Momentum Transfer Case}

\begin{figure*}[hb]
\plottwo{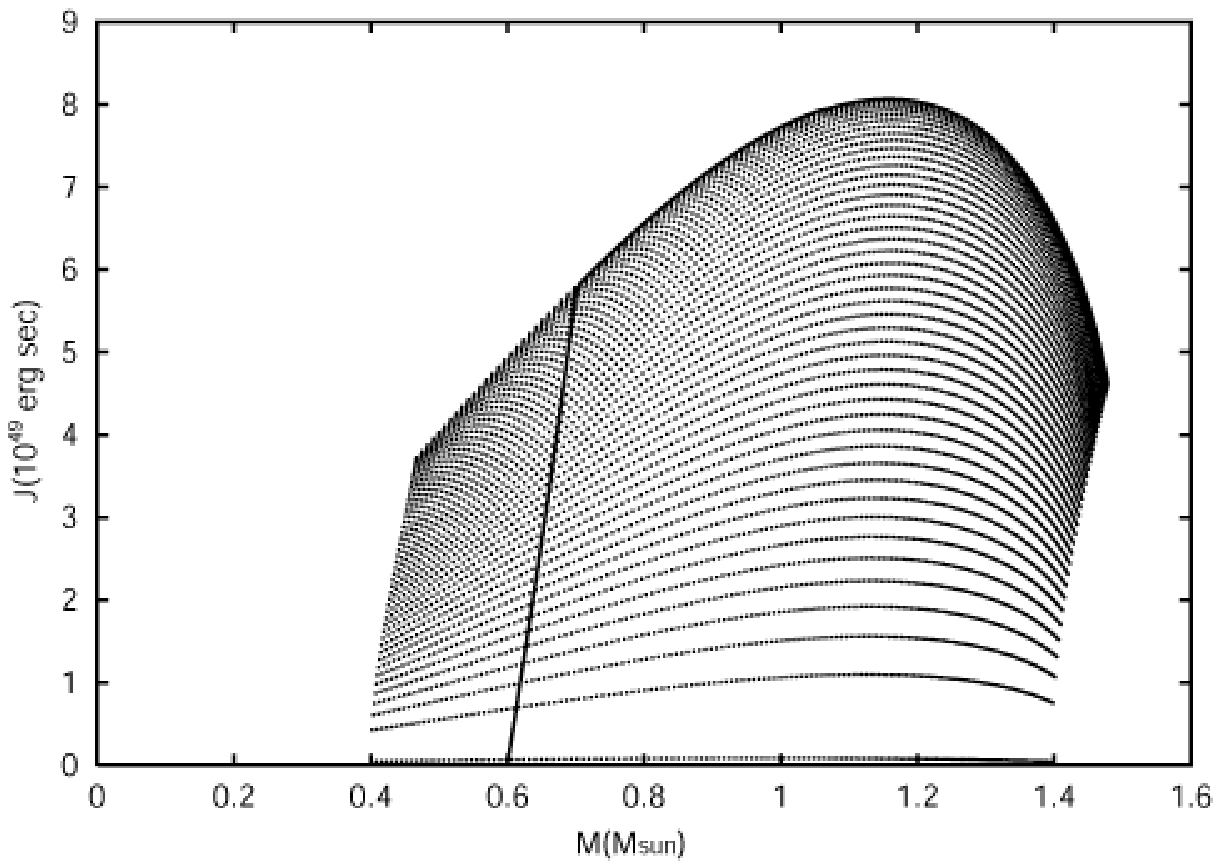}{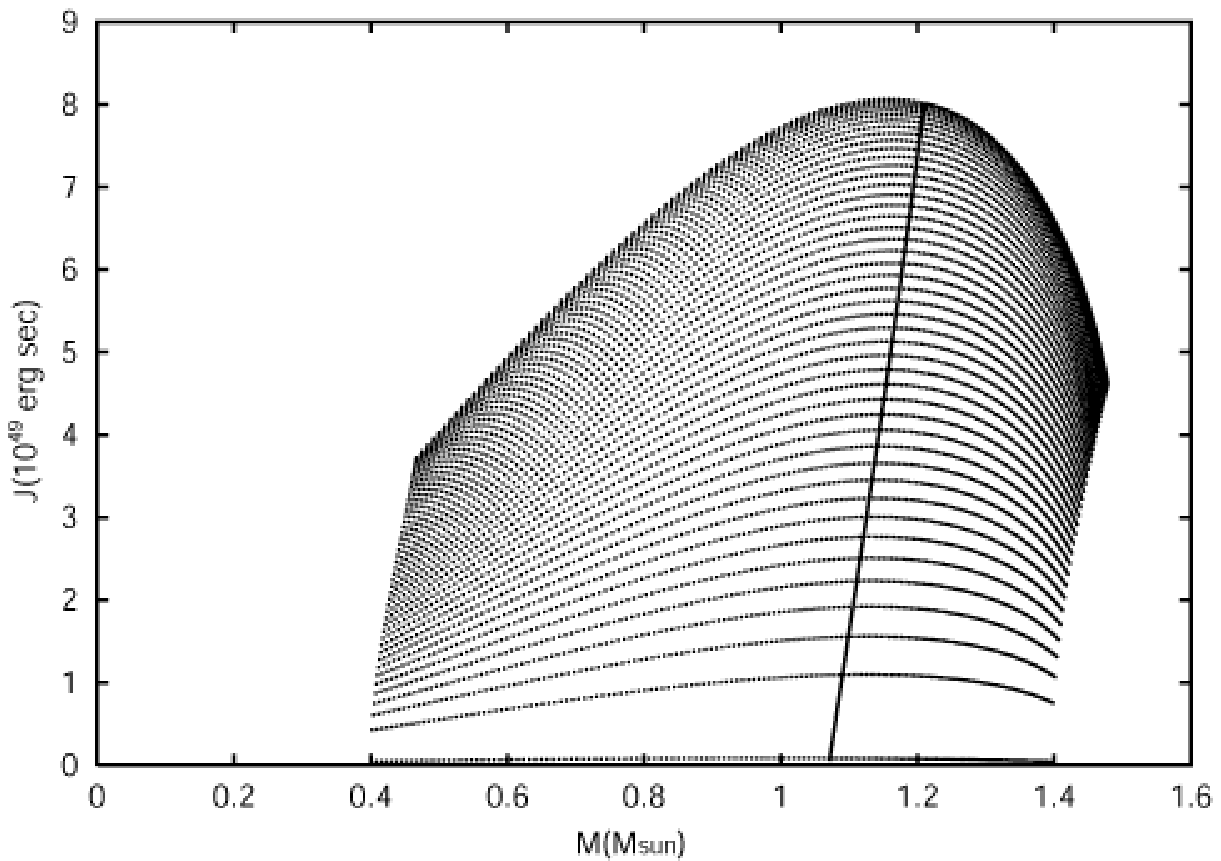}
\caption{The evolutionary tracks of accreting WDs for fast transfer case ($\xi = 0$ throughout the sequence). The left panel shows the track of $M_{\mathrm{int}} = 0.6M_{\odot}$ and the right panel shows one of $M_{\mathrm{int}} = 1.07M_{\odot}$.}
\label{Figure:Seq.fast}
\end{figure*}

Results for the fast angular momentum transfer case are shown in Figure \ref{Figure:Seq.fast}. As described in \S 2, $\xi = 0$ throughout the evolutionary sequence. The accreting WD reaches the upper edge of the data table regardless of its initial mass as seen in Figure \ref{Figure:Seq.fast}. It means that all WDs rotates fast enough to be affected by the supercritical rotation. The mass when the accreting WD reaches the supercritical rotation ($M_{\mathrm{crit}}$) varies from $0.70$ to $1.21 M_{\odot}$, as summarized in Table \ref{Table:Crit.Mass.Fast}. Although the property of WDs is diverse when they reach supercritical rotation, they will converge to a single point, which is at the upper right corner in the left panel of Figure \ref{Figure:MJ}, and explode as supernovae. The final mass $M_{\mathrm{fin}}$ is $1.48M_{\odot}$ and the final total angular momentum $J_{\mathrm{fin}}$ is $4.63 \times 10^{49}$ erg$\cdot$sec. The effect of rotation contributes to the homogenity of SNe Ia rathar than causes the diversity if angular momentum transport is much faster than accreting timescale.

\begin{center}
\begin{tabular}{ccc}
\multicolumn{3}{c}{TABLE 1}\\
\multicolumn{3}{c}{INITIAL AND CRITICAL MASS}\\
\multicolumn{3}{c}{FOR FAST TRANSFER}\\
\hline
\hline
$M_{\mathrm{int}}$ & $M_{\mathrm{crit}}$ & $J$\\
$(M_{\odot})$ & $(M_{\odot})$ & $(10^{49}$ erg$\cdot$sec)\\
\hline
0.60 & 0.701 & 5.80\\
0.70 & 0.814 & 6.70\\
0.80 & 0.925 & 7.44\\
0.90 & 1.03 & 7.90\\
1.00 & 1.14 & 8.08\\
1.07 & 1.21 & 8.03\\
\label{Table:Crit.Mass.Fast}
\end{tabular}
\end{center}

\begin{figure*}[ht]
\plottwo{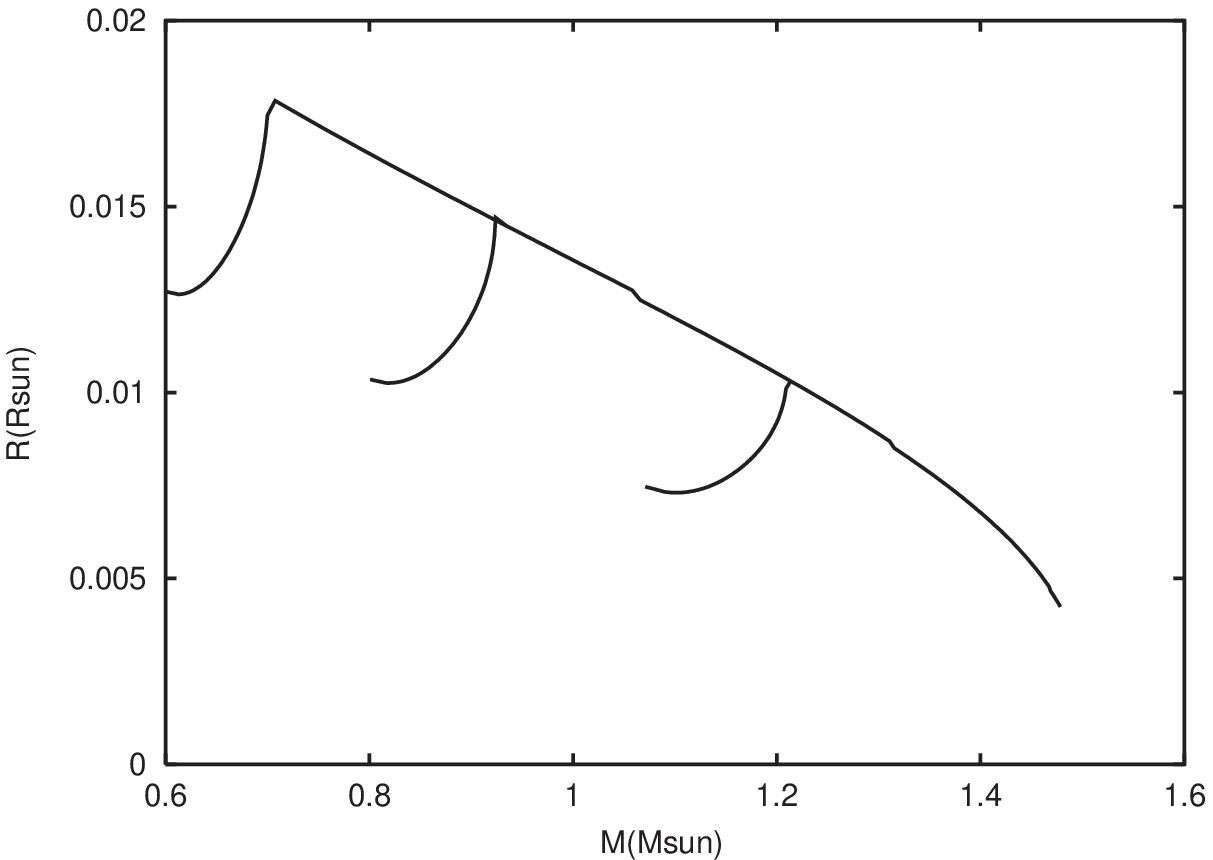}{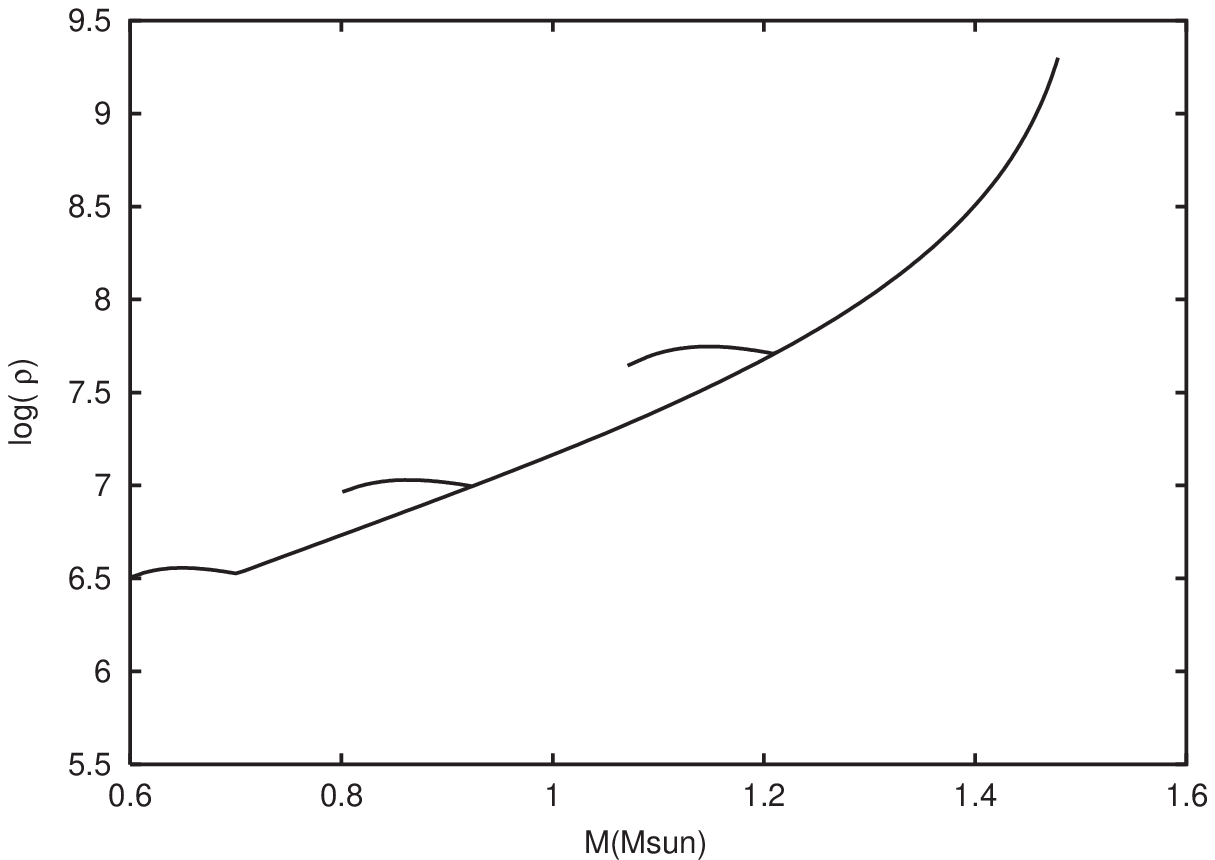}
\caption{Radius and central density of accreting WDs are plotted against mass during accretion. The left panel shows is for radius and the right panel is for central density. $M_{\mathrm{int}}$ is $0.6, 0.8, 1.07 M_{\odot}$.}
\label{Figure:R-Den.fast}
\end{figure*}

In Figure \ref{Figure:R-Den.fast}. the radius and maximum density of the accreting WD is plotted against the mass as the WD accretes matter from the companion star. Initial masses of the WDs in each sequence is $0.6, 0.8$ and $1.07 M_{\odot}$, respectively. Before reaching the critical rotation stage, radius of the WD increases because the net gravitiy is decreased as the rotation grows faster. When supercritical rotation occurs and rotation is stabilized, the radius decreases as the WD accretes matter to increase its mass. For the central density, on the other hand, the sequence is not monotonic before WDs reach supercritical rotation. Under the supercritical rotation, the central densities in all sequences converge and increase toward SNe Ia.

\subsection{Slow Angular Momentum Transfer Case}

\begin{figure*}[hb]
\plottwo{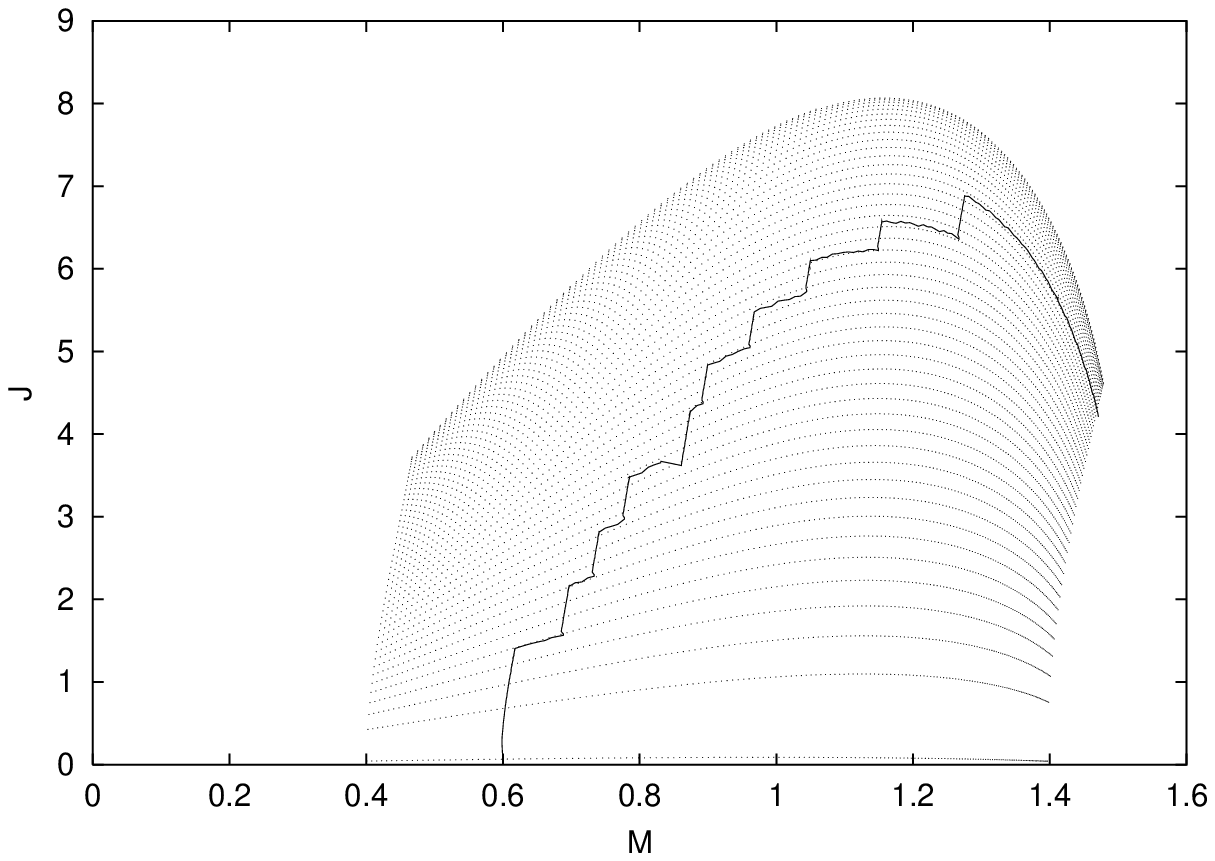}{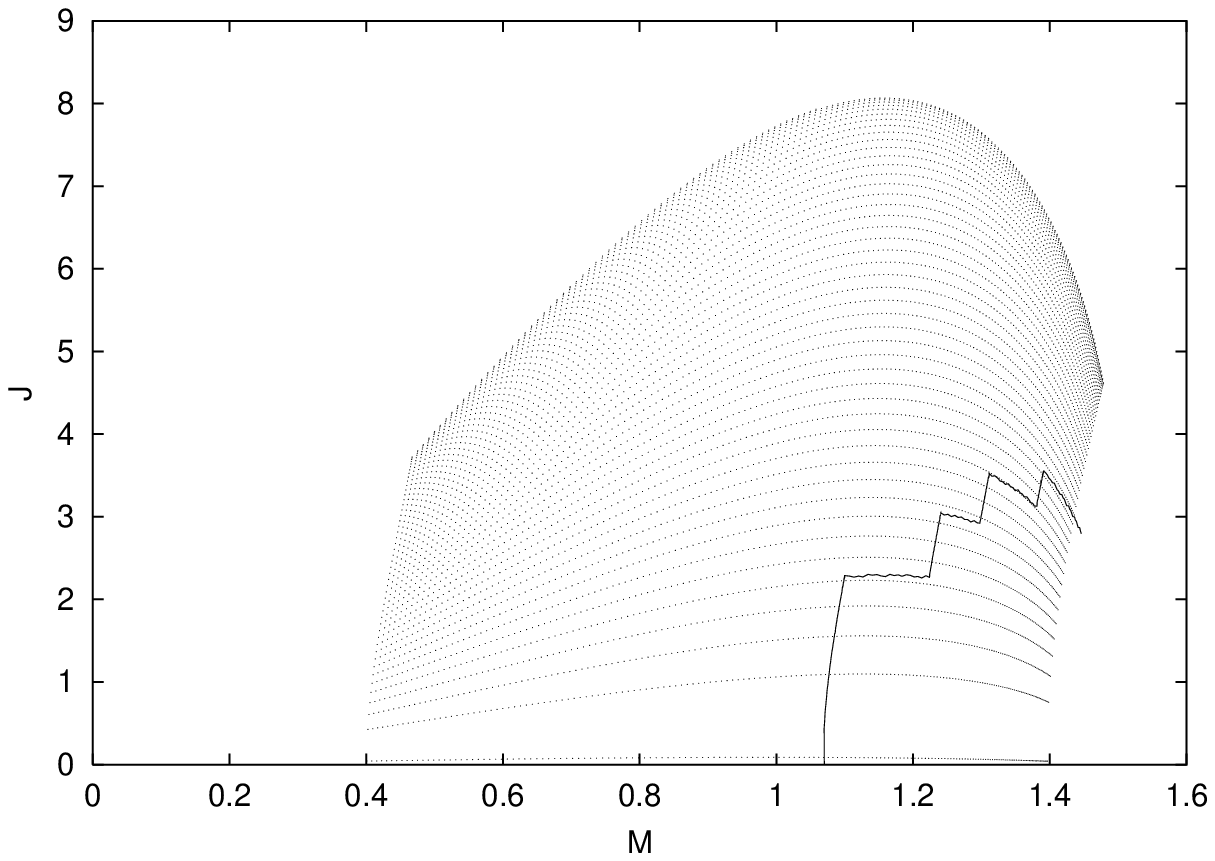}
\caption{Same as Figure 4, but for slow transfer case.}
\label{Figure:Seq.Slow}
\end{figure*}

For the slow angular momentum transfer case, results are shown in Figure \ref{Figure:Seq.Slow}. Because the value of $\xi$ changes as the WD accretes matter to increase its mass, the effect of supercritical rotation suppresses the gain of angular momentum continuously, although sequences are not smooth because of the limited resolution in parameter space of $\xi$.

Unlike the fast angular momentum transfer case, the final state of WDs are not uniform but depend on their initial mass ($M_{\mathrm{int}}$). The total mass and angular momentum of WDs just before the supernova explosion ($M_{\mathrm{fin}}, J_{\mathrm{fin}}$) are summarized in Table \ref{Table:Final.Mass.Slow}. $J_{\mathrm{fin}}$ is smaller than the fast angular momentum transfer case, because the value of $\xi$ is non-zero and the region on the $M-J$ plane where WDs can exist stably is narrower. As seen in Table $\ref{Table:Final.Mass.Slow}$, WDs with smaller $M_{\mathrm{int}}$ have larger mass and angular momentum when their central density reaches $2.0 \times 10^9 \mathrm{g/cm}^3$. The difference of mass is small ($\sim 0.02 M_{\odot}$); however, the difference of angular momentum is rather large ($\sim 2 \times 10^{49}$ erg $\cdot$ sec). 

\begin{center}
\begin{tabular}{cccc}
\multicolumn{4}{c}{TABLE 2}\\
\multicolumn{4}{c}{INITIAL MASS AND FINAL STATE}\\
\multicolumn{4}{c}{FOR SLOW TRANSFER}\\
\hline
\hline
$M_{\mathrm{int}}$ & $M_{\mathrm{fin}}$ & $J_{\mathrm{fin}}$ & $\xi_{\mathrm{fin}}$\\
$(M_{\odot})$ & $(M_{\odot})$ & $(10^{49}$ erg$\cdot$sec) &\\
\hline
0.60 & 1.47 & 4.15 & 0.42\\
0.70 & 1.47 & 3.94 & 0.49\\
0.80 & 1.46 & 3.72 & 0.56\\
0.90 & 1.46 & 3.43 & 0.63\\
1.00 & 1.45 & 3.15 & 0.69\\
1.07 & 1.45 & 2.76 & 0.75\\
\label{Table:Final.Mass.Slow}
\end{tabular}
\end{center}

\section{DISCUSSION}
We obtained the result that the accreting WDs in close binaries have the tendency that WDs with smaller initial mass ($M_{\mathrm{int}}$) have larger final mass ($M_\mathrm{fin}$) and angular momentum ($J_{\mathrm{fin}}$) when we assume that angular momentum transfer is sufficiently slow compared to the accretion rate \citep{Yoon02}.

It is reported that brightnesses of SNe Ia show the following dependence on their parent galaxies \citep{Hamuy00}.  The bright SNe Ia are observed only in spiral galaxies, while dimmer SNe Ia are found in both spiral and elliptical galaxies. Effects of rotation might qualitatively explain this tendency as follows.

In elliptical galaxies, star formation has stopped long time ago, so the massive stars have already evolved.  Therefore, a WD with $M_{\mathrm{int}}$ that is too small cannot accrete enough mass to make a supernova explosion because no companion star can supply enough matter to it \citep{Umeda99}. On the other hand, in spiral galaxies, star formation is still active, so the companion stars of WDs can be massive. Therefore, WDs of small $M_\mathrm{int}$ can get enough mass to become SNe Ia.

With the effect of rotation, accreting WDs with smaller $M_{\mathrm{int}}$ grow to larger $M_{\mathrm{fin}}$ and $J_{\mathrm{fin}}$ at the explosion. Although the difference of $M_{\mathrm{fin}}$ is small, the difference of  $J_{\mathrm{fin}}$ might affect the explosion and following nucleosynthesis. If there is a relation that $^{56}$Ni is synthesized more for larger $J$, accreting WDs with smaller $M_{\mathrm{int}}$ grow to larger $J_{\mathrm{fin}}$ and are expected to become brighter SNe Ia. This implies that the upper limit of SNe Ia brightness is higher in spirals than in elliptical galaxies. In this way, SNe Ia in spiral galaxies can originate from WDs with a wider range of $M_{\mathrm{int}}$, thus having a larger dispersion of brightness than those in elliptical galaxies. This could explain the observation that the most luminous SNe Ia appear to be observed only in spirals, while dimmer SNe Ia are observed in both spiral and elliptical galaxies.

In the above discussion we assumed that angular momentum transfer inside the accreting WDs are sufficiently slow compared to the accretion rate ($\dot{M}$), and our calculation is hydrostatic. To examine whether these assumptions are valid, more precise study using multi-dimensional hydrodynamic caluculation is needed. As an accreting WD gains matter from their companion star, the whole binary system also evolves and $\dot{M}$ should change. Therefore we have to consider not only the accreting WD itself, but also the whole binary in order to treat angular momentum transfer precisely.

Furthermore, we do not have precise theoretical knowledge of flame propagation and nucleosynthesis in the supernova explosion. In addition, the effect of rotation on these phenomenona is not clear. We have to execute multi-dimensional explosion simulations to investigate whether the rotation causes the significant diversity of the luminosities of SNe Ia.

\section{CONCLUSIONS}

We have computed the structure of the rotating accreting WDs of various axis ratio ($q$) and central density ($\rho_c$). For the rotation law, rigid rotation is assumed and the ratio of the nonrotating core to the whole WD ($\xi$) is introduced. For each value of $\xi$, region on $M-J$ plane where WDs can exist stably is obtained. When we simulate the evolutionary track of WDs in $M-J$ plane, we consider they continue to accrete matter under the effect of supercritical rotation \citep{Pac90} after reaching the critical rotation.

The effect of rotation depends on the angular momentum transfer timescale compared to the accretion rate. If angular momentum transfer is sufficiently fast, all angular momentum brought to the WD is averaged and the WD rotates rigidly. WDs reach critical rotation and converge to the single state regardless of their initial state. Therefore, rotation contributes to the uniformity of SNe Ia when angular momentum transfer is fast. On the other hand, when angular momentum transfer is slow, there exists a nonrotating central region of the WDs. The size of this nonrotating region depends on the initial mass of the accreting WDs, as the final state depends on the initial state.

We have found that for accreting WDs, a smaller initial mass $M_{\mathrm{int}}$ results in a larger final mass $M_{\mathrm{fin}}$ and angular momentum $J_{\mathrm{fin}}$ at explosion. This diversity of $M_{\mathrm{fin}}$ is about $0.02 M_{\odot}$ and may be too small to contribute to the observed diversity of brightness of SNe Ia. On the other hand, the difference in $J_{\mathrm{fin}}$ could be sufficiently large to affect the speed of flame or nucleosynthesis in the explosoion. Further study is needed to demonstrate the effect of rotation on the explosion and whether the rotation can be the source of the diversity of brighnesses of SNe Ia. 

We would like to thank Sung-Chul Yoon and Norbert Langer for useful discussion.  This work has been supported in part by the grant-in-Aid for Scientific Research (14047206, 14540223, 15204010) of the Ministry of Education, Science, Culture, Sports, and Technology in Japan.

\end{document}